# PATHOLOGICAL WATER SCIENCE – FOUR EXAMPLES AND WHAT THEY HAVE IN COMMON

Daniel C. Elton
*Radiology and Imaging Sciences, National Institutes of Health Clinical Center, Bethesda, MD 20892, USA. email : daniel.elton@nih.gov*

Peter D. Spencer
*School of Biomedical Sciences, Faculty of Health, Institute of Health and Biomedical Innovation, Queensland University of Technology (QUT), Brisbane, Australia.*

## ABSTRACT

Pathological science occurs when well-intentioned scientists spend extended time and resources studying a phenomena that isn't real. Researchers who get caught up in pathological science are usually following the scientific method and performing careful experiments, but they get tricked by nature. The study of water has had several protracted episodes of pathological science, a few of which are still ongoing. We discuss four areas of pathological water science – polywater, the Mpemba effect, Pollack's "fourth phase" of water, and the effects of static magnetic fields on water. Some common water-specific issues emerge such as the contamination and confounding of experiments with dissolved solutes and nanobubbles. General issues also emerge such as imprecision in defining what is being studied, bias towards confirmation rather than falsification, and poor standards for reproducibility. We hope this work helps researchers avoid wasting valuable time and resources pursuing pathological science.

## 1. INTRODUCTION TO PATHOLOGICAL SCIENCE

In 1953 the Nobel prize winning chemist Irving Langmuir gave a talk on pathological science, which he referred to as the "the science of things that aren't so" (Langmuir and Hall, 1989). Langmuir had observed several cases (often firsthand) where scientists were tricked into believing in a phenomena, often for years or decades. Eventually it was found the purported phenomena was actually caused by confounding factors in an experiment or faulty methods of data analysis. Some of the examples that Langmuir discussed are N-rays, mitogenic rays, and extrasensory perception. Some prominent examples since 1954 are polywater, cold fusion, and magnet therapy. Many other mini-episodes of pathological science can be found in the psychological and social sciences, which are currently undergoing a major reproducibility crisis. We believe the scientific community needs to get better at detecting pathological science. The first reason for this view is the obvious one – pathological science wastes scientist's time and (usually) taxpayer money. The second reason is that properly sorting out the "wheat from the chaff" can very hard both for other scientists and the public when great volumes of pathological science are being published. Intense competition for funding has led to rushed work and exaggerated or sensationalized findings. Increased pressure to publish has led to a proliferation of low tier journals with weaker standards of peer review. Together with the rise of preprint servers, scientists and the public now have to deal with a deluge of low-quality papers. Finally, we note that pathological science is often used to promote products which actually have no utility to the end user. This is especially a problem in the area of health-treatments because resources are sometimes misallocated away from treatments that would have actually helped the patient.

We wish to emphasize that pathological science is distinct from pseudoscience. While some pseudoscience may also be called pathological science, not all pathological science is pseudoscience. The reason not all pathological science is pseudoscience is that most researchers working on pathological science are trained career scientists who use the scientific method well. They simply are tricked! We also want to emphasize that those who have fallen prey to pathological science are generally well intentioned and often very bright and talented researchers. Even Nobel prize winners have fallen for pathological science – Brian Josephson (1973, Physics) and Luc Montagnier (2008, Physiology or Medicine) have both endorsed water memory as a real phenomenon.

The features of pathological science that Langmuir identified in his talk are:

1. "The maximum effect that is observed is produced by a causative agent of barely detectable intensity, and the magnitude of the effect is substantially independent of the intensity of the cause."
2. "The effect is of a magnitude that remains close to the limit of detectability; or, many measurements are necessary because of the very low statistical significance of the results."
3. "Claims of great accuracy."
4. "Fantastic theories contrary to experience."
5. "Criticisms are met by ad hoc excuses."

6. "Ratio of supporters to critics rises up to somewhere near 50% and then falls gradually to oblivion"

In this paper, we take a very broad view of what pathological science is. So, not every example of pathological science we discuss exhibits features 1-6. To us, pathological science is simply any area of science where nature tricks researchers into believing in a phenomenon for an extended period of time. It is our observation that research on water is particularly prone to pathological science and we explore why this might be. Liquid water, the substrate in which all known life operates, holds a privileged position in human culture and science. Phillip Ball explores this in his book *H$_2$O: A Biography of Water* and argues that the idea that "water is special" is a bias instilled in us by thousands of years of human culture (Ball, 1999). This is undoubtably true, but scientifically such a bias is not entirely off-the-mark – water does have many anomalous properties and is special in many ways amongst liquids. Issues only occur when people latch onto the idea that water is more special than it really is and then do not properly criticize their ideas and only seek confirmation of them rather than falsification. Humans are subject to many cognitive biases (Kahneman, 2011), and some of these, such as the confirmation bias and extension neglect (neglect of magnitude) undoubtably play a role in pathological science. Our eye in this work however is less on human psychology and more on the specific properties of water which make it difficult to study and thus prone to pathological science. We hope this work helps researchers studying water develop a more critical attitude and avoid wasting time pursuing pathological science.

## 2. POLYWATER

Perhaps the most famous example of pathological science is polywater. The polywater saga has been explicated in many places, so we keep are summary here brief. Polymeric water ("polywater") was purported to be a special phase of water which formed when water was condensed into tiny capillary tubes with diameters smaller than 100 micrometers. The earliest papers on polywater originated from the group of Boris Deryaguin at the Institute of Surface Chemistry in Moscow, USSR in the early 1960s. In 1962 Fedayakin proposed that polywater had a honeycomb like structure with each oxygen bonded to 3 hydrogens. Lectures by Deryaguin in England and the United

States in 1966, 1967 and 1968 drew the attention of western researchers. Research interest peaked after a 1969 a paper by Lippincott et al. in *Science* which reported spectroscopic results which were said to provide conclusive evidence of a stable polymeric structure" (Lippincott et al., 1969). Over 160 papers on polywater were published in 1970 alone (Eisenberg, 1981). In 1971 Hasted noted problems with hexagonal water structures in general, noting that high energy cost of placing hydrogens between oxygens was enough to make such structures explode if they were ever created (Hasted, 1971). By 1972 it became apparent that the observed phenomena were due to trace amounts of impurities (Rousseau and Porto, 1970), some of which likely came from human sweat (Rousseau, 1971). In some cases, it was found that the sample tubes contained very little water at all. Altogether, over 500 publications were authored on polywater between 1963-1974 (Eisenberg, 1981; Bennion and Neuton, 1976).

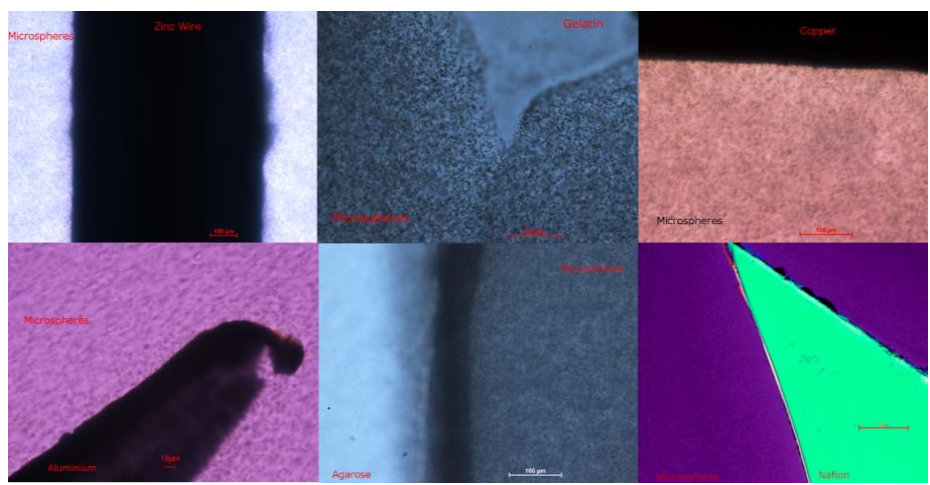

*Figure 1*. Top row from left to right – zinc, gelatin, copper. Bottom row – aluminum, agarose, and Nafion. All EZs were visualized using a 1:500 suspension of 1.0 µm carboxylated solid-latex microspheres. The last image (Nafion) is the only one which shows an exclusion zone, evidenced by the much lower density of microspheres near the surface. The image with Nafion was taken with a polarized light microscope where birefringent materials appear brightly colored.

## 3. EXCLUSION ZONE PHENOMENA AND POLLACK'S "FOURTH PHASE"

Recently we reviewed the literature on exclusion zone (EZ) phenomena in water (Elton et al., 2020). The EZ occurs when plastic microspheres are repelled away from the surface of some material leaving a region of microsphere-free water near the surface. An EZ near first observed in the laboratory of Gerald Pollack in 2003 near polyvinyl alcohol gels (Zheng and Pollack, 2003). Later, in 2006 Pollack reported larger EZs near the surface of Nafion (Zheng et al., 2006). As we review in our article, the existence of an EZ near Nafion has been replicated many times by at least 10 different laboratories and constitutes a real phenomenon in search of an explanation. The finding of an EZ near metals has only been found by two independent groups and in our own experiments we were not able to replicate it in. either zinc, copper, or aluminum with negatively charged carboxylated latex microspheres (fig. 1) (Spencer et al., 2018). It is unclear how big EZs are near hydrophilic materials other than Nafion. In his 2003 work, Pollack observed EZs near several hydrophilic gels such as polyacrylic acid, poly acrylamide, and agarose, but none of these results have been replicated. More specifically, Pollack reported that positively charged functionalized spheres were repelled by agarose gels (which is weakly negatively charged), but he does not report any results for other types of microspheres (Zheng and Pollack, 2003). In our own study we did not find an EZ near either agar, agarose, or gelatin using neutral latex microspheres (fig. 1). The pH of the agarose started very close to 7 and decreased to 6 after about 20 hours and no changes in pH were observed for agar. Despite the lack of replication of the EZ phenomena beyond Nafion, Pollack often claims in interviews that an EZ is generated near all hydrophilic materials and plays an important role in biological processes.

It appears the core phenomenon of an EZ near Nafion is real and is likely caused by diffusiophoresis (also called chemotaxis) due to a long-lived pH gradient generated by the negatively charged sulfonic groups which are particular to the surface of Nafion (Elton et al., 2020; Schurr, 2013; Florea et al., 2014b; Musa et al., 2013). Functionalized microspheres contain surface charges which lead to counterions near their surface. The surface charge is distributed uniformly, but a non-uniform pH gradient causes the counter ion distribution to become non-uniform. This sets up different electrostatic forces on both sides of the particle, leading to a net force on the particle (figure 2 illustrates this). As shown by Florea, the theory of diffusiophoresis precisely explains the kinetics (growth) of the EZ over time. The existence of a large pH gradient near Nafion has been shown using indicator dye in several of Pollack's works (for instance (Chai et al., 2009b)).

Indeed, it seems the EZ phenomenon isn't really specific to water – research from Pollack's own lab showed that it occurs in a variety of liquids - methanol, ethanol, isopropanol, acetic acid, and dimethyl sulfoxide (Chai and Pollack, 2010). Further investigations into EZ water, however, have generated much work we regard as verging on pathological because it meets several of Langmuir's criteria (in particular principles 1, 3, and 4). While Pollack is usually careful about what he says in his journal articles, typically sticking to the observed experimental facts, Pollack's book contains many wild conjectures which fly in the face of basic science, such as the idea that bloodflow is powered by sunlight (Pollack, 2013). The most famous of these is Pollack's proposal that the EZ contains a "fourth phase" of water – a claim which is explored and discussed in several of his peer reviewed papers as well (Chai et al., 2009a). Pollack hypothesizes that EZ water is structured in hexagonal sheets, with the hydrogens lying directly between oxygens, a structure which is very similar to polywater. He further proposes that when these sheets are stacked, hydrogen atoms bond to the oxygens in neighboring layers such that each hydrogen forms three bonds. Oehr and LeMay present a similar theory that the observed EZ water may comprise tetrahedral oxy-subhydride structures (Oehr and LeMay, 2014). Pollack also hypothesizes that when light is shined on EZ water it causes positive and negative charges to separate, and the EZ water region to grow (Chai et al., 2009b). This is obviously problematic since water is a good conductor and charge separation would be difficult to sustain.

Pollack points to enhanced absorption at 270 nm as evidence for a possible phase change in the EZ (Zheng et al., 2006; So et al., 2012). This absorption peak was not found in quantum chemistry simulations (Segarra-Martí et al., 2014). Strikingly, results from Pollack's own lab show that a similar absorption peak is seen in pure salt solutions (LiCl, NaCl, KCl) (Chai et al., 2008), so the source of this enhanced absorption appears to be related to dissolved solutes. A study of Arrowhead Spring water found absorption at 270 nm, so even trace dissolved solutes can create it (Dibble et al.). Hypothesizing that EZ water would be a transitionary form between ice and liquid water, Pollack performed UV absorption measurements of melting ice (So et al., 2012). During the course of these experiments the 270 nm peak sometimes (but not always) appeared transiently (i.e., for a few seconds) while the ice was melting (Langmuir's criteria #1 and #2) . In the same work they also report that degassing the water reduced of the appearance of the peak (So et al., 2012). Thus, it is also possible that the peak is related to tiny bubbles trapped in the ice which migrate to the surface while the ice is melting. As we discuss in our review (Elton et al., 2020), a possible mechanism for the absorption near 270 nm would absorption from superoxide anions ($O_2^-$) and their protonated form, the hydroperoxyl radical ($HO_2$). Such absorption may be enhanced by nanobubbles.

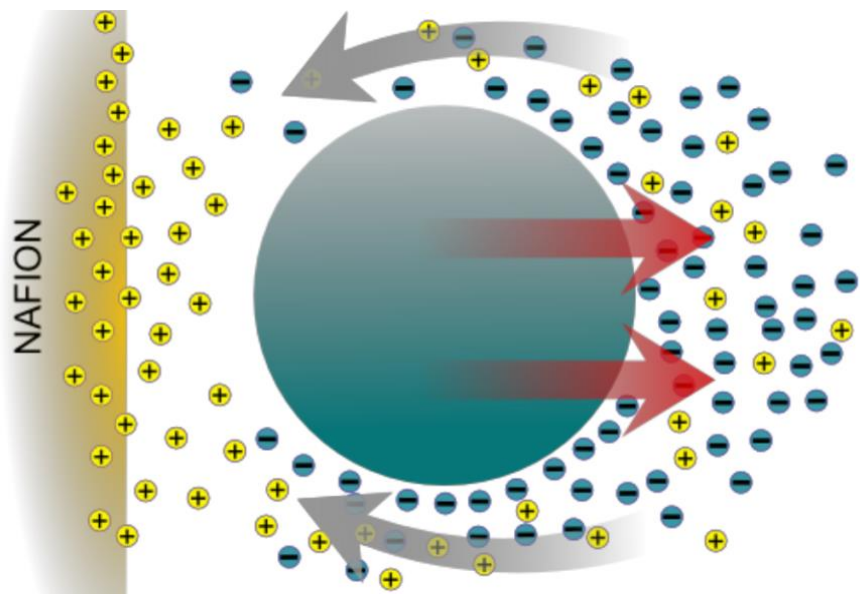

*Figure 2*. Illustration of the mechanism by which diffusiophoresis generates a force on plastic microspheres which leads to the exclusion zone (Schurr, 2013; Florea et al., 2014a).

Pollack's promotion of his fourth phase theory deserves to be vigorously criticized not only because it contradicts basic thermodynamics, but also because it lends support to a sprawling number of enterprises selling "structured" or "hexagonal" water for health purposes. Tests of some of these products with nuclear magnetic resonance spectroscopy (NMR) show no difference from pure water (Shin, 2006). Companies currently selling EZ water products who cite Pollack's work include Divinia Water, Structured Water Unit LLC, Flaska, Advanced Health Technologies (vibrancywater.ca), and Adya Inc. The idea of utilizing EZ water for health has also been promoted by influential figures in alternative medicine such as Dr. Joseph Mercola and Dave Asprey. Instead of providing much needed words of scientific skepticism caution, Pollack has embraced the attention he has received from alternative medicine community by participating in podcasts with Mercola, Asprey, and many others where he has promoted the idea that EZ water is important for health.

## 4. THE MPEMBA EFFECT

The idea that hot water can freeze faster than cold water has a long history (Jeng, 2006). Brief mentions of this phenomena can be found in the writings of several famous thinkers including Aristotle, Thomas Bacon, and Descartes (Jeng, 2006). In 1969 a Tanzanian high school student named Erasto Mpemba co-authored an article on the subject in the journal *Physics Education* (Mpemba and Osborne, 1969). Erasto was actually studying sugared milk while making ice cream, but his finding spurred research searching for the effect in pure liquid water. In the period of 1970-1990, dozens of papers were published which purported to find such an effect. The literature is confusing due to the lack of experimental standards leading to many variables coming into play (some used distilled water, some used tap, some studied the effects of dissolved salts, authors used different cooling schedules/methods, etc). It appears none of the studies attempted to explicitly replicate a prior study, which resulted in all the studies having different types of experimental setup. Researchers also used slightly different definitions regarding the precise circumstances that would constitute confirmation or falsification of the Mpemba effect. Katz (2008) has analyzed this perplexing literature and postulates that most of the experiments were contaminated by solutes, either gaseous or solid (Katz, 2008). He proposes that dissolved solutes (either gaseous or solid) are removed during heating, and that solutes accumulate along the freezing front and reduce the heat flux. Later, Linden & Burridge also reviewed the prior literature (Burridge and Linden, 2016). They also performed their own study which showed that the height at which the temperature is measured determines what relative freezing times are observed. Since most prior work did not report this variable, it is hard to compare literature results. The conclusion from their own experiments was that the effect does not exist (Burridge and Linden, 2016).

`

Assuring that containers with hot and cold water are cooled in identical and measured in an identical fashion which accurately determines the freezing time requires a careful experimental setup. To give a simple example of a pitfall that students might encounter at home, freezers have a thin layer of ice crystals coating their interior surfaces. If you place a container of hot water in such a freezer, the ice crystals will melt, allowing for better thermal contact between the container and the freezer. Thus, it's not surprising the container with the hot water freezes faster in such case.

It has only been recently that very careful experiments have been performed which attempt to cool hot and cold water under perfectly identical conditions. One such series of experiments was published by Dr. James D. Brownridge in 2010 (Brownridge, 2010). Brownridge took ultra-pure samples of distilled water, sealed them in small glass vials and suspended the vials by threads in a vacuum. The vials were then cooled using radiative cooling. This completely removed the possibility of a difference in the thermal contact between the hot & cold vials and ensured they were cooled in exactly the same fashion. Brownridge found that in some cases that the hot water vial would freeze first. This only occurred though when the cold water would supercool further than the hot before freezing. At all times, the hot water was always warmer than the cold water, and both vials were cooling at the same rate – it was just the cold water supercooled more. Brownridge found that each glass vial has a highest temperature nucleation site (HTNS) which determines the temperature water will freeze in that vial (Brownridge, 2010). Comparing what appear to be identical vials, the HTNS are random and they can be between anywhere from 0 C to -45 C. Brownridge showed that the HTNS is a constant of the container by rerunning the freezing many times. So, in the end, the two containers (hot and cold) were not actually identical because they had different nucleation sites! The basic idea behind this -- that supercooling to different degrees as a result of unpredictable nucleation factors was responsible for the Mpemba effect being observed, had been proposed earlier by Auerbach in 1995. (Auerbach, 1995).

Brownridge and Auerbach's work showing that the Mpemba effect is just due to unpredictable supercooling seems to have been completely lost on the Royal Society of Chemistry, which in 2012 held a much-publicized competition to explain the effect. The winner of that competition proposed that the effect is due to some or all of the following: (a) evaporation, (b) dissolved gases, (c) mixing by convective currents, and (d) supercooling. Brownridge, by carefully removing the confounds of (a)-(c), showed that supercooling is enough to generate the effect if containers with different nucleation sites are used. Still, (a) – (c) are quite possibly confounding factors which were responsible for observations of the Mpemba effects in previous works.

## 5. MAGNETIC FIELDS AND WATER

The number of different effects that magnets have been claimed to have on water is truly mind boggling, and there are too many to properly analyze in this small chapter. Magnetic fields have been reported to change the physicochemical properties of liquid water, including viscosity (Ghauri and Ansari, 2006; Cai et al., 2009), refractive index (Hosoda et al., 2004), melting temperature (Inaba et al., 2004), rate of vaporization (Nakagawa et al., 1999), adsorption (Ozeki et al., 1991; Higashitani et al., 1993), electrolyte conductivity (Holysz et al., 2007), and conductivity (Szcześ et al., 2011). Some authors report that the property changes remain for many hours even after the magnetic fields are turned off (Mahmoud et al., 2016; Silva et al., 2015; Coey and Cass, 2000; Szcześ et al., 2011). Magnetic fields have been claimed to inhibit the formation of ice crystals in both pure water and biological products (Otero et al., 2016). There is also research that purports that magnetic fields can be used to "treat" water in some way – either to purify (Ambashta and Sillanpää, 2010), de-scale (Coey and Cass, 2000), or disinfect water (Biryukov et al., 2005). Authors who have attempted to review this massive and perplexing literature have lamented the lack of independent reproduction of most results (Knez and Pohar, 2005; Smothers, 2001). Some of these experimental findings are said to be supported by molecular dynamics simulations that show that magnetic fields enhance hydrogen bonding (Chang and Weng, 2006). However, the effect size is extremely small – a 10 Telsa magnetic field caused a size increase of only 0.34% in water clusters in one such study(Toledo et al., 2008). This is not surprising because the magnetic susceptibility of water molecules is very small – about $-9.0 \times 10^{-6}$ (Otero et al., 2016).

Research on the effect of magnetic fields on the freezing of water is very mixed. For instance, work in 2000 on pure water droplets found that magnetic fields *reduced* the degree of supercooling before freezing in contrast to many works which suggest that magnetic fields inhibit freezing (Aleksandrov et al., 2000). That is not to say that magnetic fields have no effect – water is weakly diamagnetic so strong enough magnetic fields induce a magnetic dipole in the opposite direction. One group of researchers levitated water droplets in a 15 T magnetic field and found that the droplets supercooled to -10° C (Tagami et al., 1999). This not a remarkable degree of supercooling, especially for droplets of pure water not in contact with any nucleating agents. There are also papers on the effects of "oscillating magnetic fields" on water but, as any physicist knows, oscillating magnetic fields are always accompanied by oscillating electric fields, so speaking of them in isolation doesn't make much sense. To the degree that a weak oscillating magnetic fields inhibits freezing in food, as has been claimed by the Japanese company ABI Corporation with their "Cells Alive" freezer, this is may be due to heating of trace metals (iron etc.) in the food, or to non-magnetic sources altogether such as acoustic vibrations in the freezer system (Wowk, 2012).

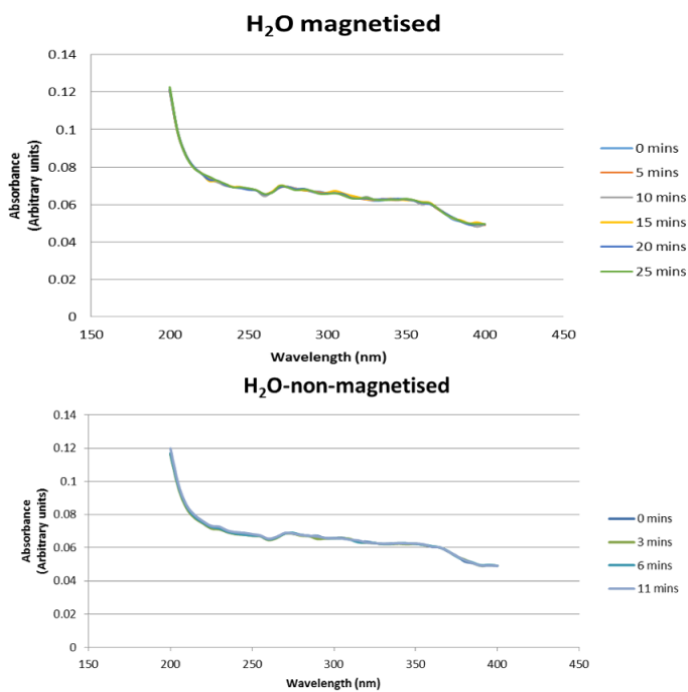

*Figure 3.* Measurements from a simple experiment one of the authors (Peter D. Spencer) performed which looked at the UV-vis absorbance of water under a magnetic field and no magnetic field (Spencer, 2018). The magnetic field was weak (0.63 T) but typical of the field strength found in studies of magnetic water treatment systems.

An analysis of the massive literature on magnetic fields and water could easily fill several review articles so in the remainder of this section we focus on a specific subfield - magnetic treatment for preventing scaling/corrosion. There is no universally agreed upon mechanism by which magnetic fields inhibit scaling but a common theory is that they work by changing the morphology of the precipitates to prevent them from depositing in flat sheets (Barrett and Parsons, 1998; Gehr et al., 1995; Holysz et al., 2007; Madsen, 1995; Higashitani et al., 1993). A review by Baker and Judd investigates numerous claims on this matter (1996). Their view is that contamination effects are the main contributor and therefor the results obtained in some experiments will not generalize to more general situations as has been claimed. In particular, they note several experiments where magnetically-enhanced corrosion likely created $Fe^{2+}$ ions which are known to retard the growth rate of calcite scale deposition. They also note that more successful results are obtained with magnetic fields orientated orthogonally to flowing water within recirculated systems. This implies that Lorentz forces acts on particulates in the water rather than the water itself. That is, forces are exerted on charged particles passing through the magnetic field.

Differences in infrared and Raman spectra in magnetically treated water have been interpreted as implying the development of quasi-stable water clusters in a magnetic field which somehow persist after the magnetic field is switched off (a variant of the "water memory" idea). This is very hard to believe given that hydrogen bond lifetimes are around 1ps in room temperature water and the Debye relaxation time is ~8-9 ps (Elton, 2017). Interestingly, Ozeki et al. found that the effect of magnetic treatment on IR absorption increases with increased dissolved oxygen and water which was fully degassed does not show any changes after treatment(Ozeki and Otsuka, 2006). They theorize that magnetic treatment leads to the formation of oxygen clathrate-like hydrates which influence the H-bond network of water (Ozeki and Otsuka, 2006). Additionally, both Lee et al.(2013) and Szcześ et al. (2011) have also reported that the concentration of dissolved gases significantly affected their results. One of us, (Peter D. Spencer) performed a simple experiment which found no effect of a magnetic field on UV-vis absorption (fig. 3) (Spencer, 2018). The field was weak (0.63 T) but typical of the field strength employed in many studies of magnetic water treatment.

# COMMONALITIES AND CONCLUDING THOUGHTS

In this work we reviewed four areas of pathological science. Due to time and space limitations we did not discuss another major area – water memory. The interested reader can consult the chapter in this very book by Yuvan and Bier which discusses it in some detail. Additionally, much water memory research, in our view, often crosses out of the domain of pathological science into pseudoscience. Water memory research has its origin in *Nature* paper from 1988 and has been thoroughly debunked (Maddox et al., 1988). Still, much work continues on water memory partially because it is a mechanism those working in the lucrative homeopathy industry have latched onto as a means of scientifically justifying their work. Homeopathy has been thoroughly debunked many times and many places (for instance a metareview of metareviews found no effect (Ernst, 2002)).

In the course of this work we have noticed a few different common features of pathological water science. The main one is improper removal of confounding factors. It is very difficult to remove dissolved solutes from water, and work indicates they were responsible for most of the experimental results in the four areas explored here. More research is needed on nanobubbles which are a possible confound and are very hard to remove from water (Jadhav and Barigou, 2020; Ball, 2012; Michailidi et al., 2020). Referring to microsphere suspensions, Horinek et al. note "these systems are notoriously plagued by secondary effects, such as bubble adsorption and cavitation effects or compositional rearrangements" (Horinek et al., 2008). Dissolved gases (not bubbles) can also be a confound as was seen in the spectral analysis of EZ water. To give another example, in 2010 Jansson et al. measured the dielectric function of water at very low frequencies and reported an "ultra-slow" Debye relaxation at 5 MHz (Jansson et al., 2010). Later work has indicated that this peak was due to microscopic bubbles in the liquid (Richert et al., 2011). Alternatively, it has been suggested that the low frequency peak is due to volatile non-polar contaminants (Casalini and Roland, 2011). It is possible both mechanisms were at play in Casalini & Roland's experiment since they observed two ultra-low frequency Debye peaks.

Another thing we noticed is lack of precision in defining the phenomena being measured. In postmodernist literature and other fields a misleading and faulty type of argument called the "motte and baily" fallacy has been identified (Boudry and Braeckman, 2011). In the "motte and bailey" style of argument, a proponent argues for a strong claim but then retreats to a much weaker claim under pressure from counterarguments. The weaker claim is then later conflated with the stronger one, sowing confusion and putting critics in a difficult position. For instance, researchers may proclaim their work shows "structure change in cellular water" or that "hot water freezes faster than cold" but under pressure will retreat to a weaker and much less interesting claim such as "proteins can reorient waters near their surface affecting 1-3 layers of water" or "hot water sometimes is observed to supercool more than cold water". We suggest that researchers focus on developing precision in their statements about experimental measurements and what they show, with a particular focus on *effect size*, which is often omitted by popular press coverage of research and thus misleads the public in unhealthy ways.

Researchers should also make sure their work is reproducible by explaining how the experiment was carried out, utilizing supplementary information if necessary to list all relevant details. Ideally, all raw data generated should be made publicly available so that the data analysis methods employed can also be reproduced. Making a full description of experimental methods and the raw data available also helps other researchers identify errors and questionable research practices (Gadomski et al., 2017).

Finally, we suggest that all researchers should evaluate their "evidence threshold" to immunize themselves from falling into believing in pathological science. One's evidence threshold is the threshold needed to believe that a proposed phenomenon is real. Dr. Steven Novella, author of the blog *Science Based Medicine,* has suggested four criteria for a good evidence threshold, the statement of which we believe is a fitting way to conclude this chapter (Novella, 2013):

1 - *"Methodologically rigorous, properly blinded, and sufficiently powered studies that adequately define and control for all relevant variables (confirmed by surviving peer-review and post-publication analysis)."*
2 - *"Positive results that are statistically significant."*
3 - *"A reasonable signal to noise ratio."*
4 - *"Independently reproducible (and reproduced). No matter who repeats the experiment, the effect is reliably detected."*

# ACKNOWLEDGEMENTS

Daniel C. Elton contributed this article in his personal capacity. The opinions expressed in this article are the author's own and do not reflect the view of the National Institutes of Health, the Department of Health and Human Services, or the United States government. Peter D. Spencer was partially supported by the Research Training Program (Stipend) funded by Department of Education and Training (Australia).